\documentclass[12pt]{article}

\input epsf
 \usepackage{amssymb,wrapfig}
\usepackage{xypic,multirow}
\usepackage[matrix,arrow,curve]{xy}
\usepackage{cite}

\makeatletter
\@addtoreset{equation}{section}
\makeatother

\begin{document}

\addtolength{\baselineskip}{.5mm}
\newlength{\extraspace}
\setlength{\extraspace}{1.5mm}
\newlength{\extraspaces}
\setlength{\extraspaces}{2mm}

\newcommand{\bz}{{\bar z}}
\newcommand{\RR}{{\mathbb R}}
\newcommand{\CC}{{\mathbb C}}

\def\nn{{\cal N}}
\def\rr {{\Bbb R}}
\def\cc {{\Bbb C}}
\def\pp {{\Bbb P}}
\def\zz {{\Bbb Z}}
\def\del {\partial}
\def\cy {Calabi--Yau}
\def\ka {K\"ahler}
\def\de {\partial}
\def\vol {\mathrm{vol}}
\def\sla#1{\rlap{\begin{picture}(10,10)
\put(0,0){\line(1,1){10}}
\end{picture} }#1}
\def\slas#1{\rlap{\begin{picture}(10,10)(-5,0)
\put(0,0){\line(2,1){15}}
\end{picture}} #1 }
\def\slash#1{\rlap{\begin{picture}(10,10)
\put(0,0){\line(5,1){40}}
\end{picture}}#1}
\def\nn{\nonumber}
\newcommand{\inv}[1]{{#1}^{-1}} 
\def\+{{+\!\!\!+}}

\newcommand{\Section}[1]{\section{#1} \setcounter{equation}{0}}

\begin{titlepage}
\begin{center}

                                \hfill   hep-th/0610210\\
                                \hfill   SU-ITP-06/28\\
                                \hfill   CALT-68-2612\\
\vskip .3in \noindent


{\Large \bf{The general (2,2) gauged sigma model \\\vskip .1cm
with three--form flux}}

\bigskip

 Anton Kapustin$^a$ and Alessandro Tomasiello$^b$\\

\bigskip

$^a$California Institute of Technology,
Pasadena CA 91125, USA \\
$^b$ITP, Stanford University, Stanford CA 94305-4060, USA


\vskip .5in
{\bf ABSTRACT }
\vskip .1in
\end{center}

\noindent We find the conditions under which a Riemannian manifold
equipped with a closed three-form and a vector field define an
on--shell $\mathcal{N}=(2,2)$ supersymmetric gauged sigma model.
The conditions are that the manifold admits a twisted generalized
K\"ahler structure, that the vector field preserves this
structure, and that a so--called generalized moment map exists for
it. By a theorem in generalized complex geometry, these conditions
imply that the quotient is again a twisted generalized K\"ahler
manifold; this is in perfect agreement with expectations from the
renormalization group flow. This method can produce new
$\mathcal{N}=(2,2)$ models with NS flux, extending the usual
K\"ahler quotient construction based on K\"ahler gauged sigma
models.

\vfill
\eject


\end{titlepage}

\section{Introduction}

The geometry of flux backgrounds in string theory has recently
become clearer. For example, the geometry underlying the most
general (2,2) nonlinear sigma model has been long described as
``bihermitian''\cite{ghr}. It describes what happens when the
complex structures $I_\pm$ felt by the left-- and right--movers
are allowed to be different, and, crucially, when a Neveu-Schwarz
(NS) three-form $H$ is introduced. For $I_+=I_-$ and $H=0$ the
manifold will be K\"ahler as familiar; but in general, for $H\neq
0$, the manifold will not be K\"ahler with respect to either
complex structure. This leads to some loss of computational power.

Twenty years after its definition, however, bihermitian geometry
has been reinterpreted as generalized K\"ahler geometry
\cite{gualtieri}.\footnote{What we call generalized K\"ahler
geometry in this paper is called twisted generalized K\"ahler
geometry in \cite{gualtieri}.} Among the applications of the new
approach are new constructions of (2,2) sigma models and an
expression for the off--shell action (extending the usual $\int
d^4 \theta K$ for the K\"ahler case) in the case in which $I_+$
and $I_-$ do not commute \cite{lrvz}. It has also been applied to
topological sigma models (for example \cite{kapustin,kl,pestun})
and to
 $\mathcal{N}=2$ NS vacua in supergravity \cite{jw}. The broader field of
generalized {\it complex} geometry \cite{hitchin,gualtieri} has
also led to a similar classification of Ramond-Ramond
$\mathcal{N}=1$ vacua \cite{gmpt2} and to related developments
concerning brane calibrations \cite{ms,martucci}.

Most of these developments are formal; it would be nice to have
tools to produce explicit examples of flux compactifications. In
the case without fluxes one such a tool is the gauged linear sigma
model \cite{w-phases}, which leads to the so--called K\"ahler
quotient construction -- a particular case of hamiltonian
reduction \cite{mw}.

It is the aim of this paper to generalize this useful tool to the
case with $H$-flux. Specifically, we work out the conditions for
$(2,2)$ supersymmetry of the general gauged sigma model.

As we have mentioned, the most general (2,2) ungauged sigma model
was written in \cite{ghr}. However, unless the two complex
structures commute, the off--shell formulation of this model
requires the introduction of complicated semi--chiral multiplets.
For this reason, the (2,2) gauged sigma model was analyzed in
\cite{hps} only in the case $[I_+,I_-]=0$, and the general case
was left alone.

While at that time this omission was justifiably perceived as that
of a pathological case, new developments have changed the
situation somewhat. As an example, in the more general setting of
Ramond-Ramond (RR) vacua, cases with non--commuting complex
structures are relevant for example for supergravity duals to
superconformal theories (as recently demonstrated for example in
\cite{mpz} for the beta--deformation in \cite{lm} and pointed out
on general grounds in \cite{wijnholt} and \cite{martucci}). One
can expect the generalized K\"ahler case to provide a stepping
stone towards finding such solutions\cite{ht}, much as the process
of adding branes on the tip of a conical Calabi--Yau gives
Sasaki--Einstein gravity duals.

We perform our analysis on shell, even though the problem of
finding the general off--shell action in the non--commuting case
has now been solved in \cite{lrvz}, as mentioned above. The
off--shell approach to gauging $(2,2)$ sigma-models has been
explored in the very recent paper \cite{mpzv}. We first perform
the physical computation in section \ref{phys}. Although the
result is formally identical to the commuting case, the
computation is lengthy, and details are left to the appendix
\ref{app:boring}. We then proceed to interpret the result
geometrically in section \ref{geometry}. In this latter task our
work is facilitated by a series of papers that appeared last year
\cite{lt1,lt2,bcg,hu,sx,vaisman} which analyzed the conditions for
generalized complex and generalized K\"ahler reduction. The
motivation for those papers was mathematically clear: given that
generalized complex geometry has symplectic geometry as an
important special case, and that hamiltonian reduction is an
important result in symplectic geometry, it is natural to wonder
if a generalized complex (or K\"ahler) structure can be reduced
too.

We refer in particular to \cite{lt2}, which contains a theorem
whose hypotheses are exactly the same as the conditions we find
for the (2,2) gauged sigma model. We find, then, perfect agreement
between that mathematical theorem and physical expectations.

Another advantage of having such mathematical literature available
(one which in fact constituted a major motivation for this work)
is the possibility to tap into the reservoir of explicit examples
those papers contain. It turns out, unfortunately, many of them
are unsuitable for physical consideration, for reasons we discuss;
we do, however, provide illustrations for several of the general
features. In particular, we provide an example with flux which is
a one-parameter deformation of the standard bihermitian structure
on $S^3\times\RR$; from the physical viewpoint, it is a $(2,2)$
deformation of the near-horizon geometry of the NS five-brane in
flat space. We also sketch an example of reduction without flux
but with non--commuting complex structures.

One possible direction in which it would be interesting to
extend the present paper is the addition of a potential term to the
sigma model. Similarly to \cite{w-phases}, this should produce
more interesting examples, presumably not yet considered in the
mathematical literature.

\section{The (2,2) gauged sigma model}
\label{phys} In this section, we describe the gauged sigma model
in terms of (1,1) superfields. In the following section we will
show how this is connected to the existence of a moment map. We
will start, however, with a subsection showing how already the
{\it ungauged sigma model} implies the existence of two moment
maps. We anticipate that the condition for the existence of the
{\it gauged} sigma model will be stronger, in that these two
moment maps are to be equal.

\subsection{The supersymmetric ungauged sigma model and two moment maps}
\label{sec:ungauged} We start from the ungauged (1,1) sigma model
whose target is a Riemannian manifold $M$ equipped with a closed
three-form $H$:
\begin{equation}
    \label{eq:ungauged}
    - \int_\Sigma  g_{mn} D_+ \phi^m D_- \phi^n \,d^2x d^2\theta
+\int_B   H_{mnp} \del_t \phi^m D_+ \phi^n  D_-\phi^p\,d^3x d^2\theta
\end{equation}
where the three-manifold $B$ is such that $\del B= \Sigma$, and
$t$ is a local coordinate in $B$ normal to $\Sigma$, such that
$\frac{\partial}{\partial t}\vert_\Sigma$ is an outward normal. If
one imposes an extended supersymmetry
\begin{equation}
    \label{eq:ung-deltaphi}
\delta \phi^m= \epsilon^+ D_+\phi^n I^m_{+\,n}  +
\epsilon^- D_-\phi^n I^m_{-\,n} \ .
\end{equation}
one gets \cite{ghr} that $I_\pm$ are complex structures, that the fundamental forms
$\omega^\pm_{mn}\equiv g_{mp}I^p_{\pm\,n}$ are antisymmetric, and that
\begin{equation}
    \label{eq:susy}
    \nabla_\pm \omega_\pm = 0 \ , \qquad ( \Rightarrow
d\omega_\pm= \pm\iota_{I_\pm} H)\
\end{equation}
where $\nabla_\pm$ are covariant derivatives with connection
$\Gamma^{LC}\pm\frac12 g^{-1}H$.
(This computation is a particular case of the one we perform later on for the
gauged model, so we do not review it here.)
This geometry is called {\it bihermitian geometry} because the forms $\omega_\pm$
make the geometry hermitian in two ways; it has been studied by
mathematicians (see for example \cite{agg}) and then shown to be equivalent
to generalized K\"ahler
geometry (to be reviewed in section \ref{geometry}) in \cite{gualtieri}.

If one further imposes invariance of the model under a
one-parameter family of diffeomorphisms generated by a vector
field $\xi$, one also gets \cite{hps}
\begin{equation}
    \label{eq:Lxi}
L_\xi H= L_\xi I_+= L_\xi \omega_+=0   \ .
\end{equation}
In particular it follows that (locally)
\begin{equation}
    \label{eq:alpha}
    \iota_\xi H= d\alpha_\xi
\end{equation}a
for $\alpha_\xi$ some one--form.

Introduce now, temporarily, the operator $\del$ and the $(p,q)$ form
decomposition for $I_+$.
 Following \cite{gpp,hps}, rewrite $L_\xi I_+=0$ as
\begin{equation}
    \label{eq:delxi}
[\del, \iota_{\xi^{0,1}}]=0\ ;
\end{equation}
taking the $(1,0)$ part of (\ref{eq:susy}) and using
(\ref{eq:delxi}), one gets $\del( \iota_{\xi^{0,1}}\omega + i
\alpha^{1,0})=0$, which locally implies
\begin{equation}
    \label{eq:del}
    \iota_{\xi^{0,1}}\omega + i \alpha^{1,0}= \del f
\end{equation}
for some function $f$. Summing this with its complex conjugate, and
using that on a function $d^c=\iota_{I_+} d$, one gets
\begin{equation}
    \label{eq:momentum}
    \iota_\xi \omega + \iota_{I_+} (\alpha + d({\rm Im} f)) = d{\rm Re} f\ .
\end{equation}
We can now notice that $\alpha$ is only defined up to exact forms anyway
(see (\ref{eq:alpha})), and reabsorb $d{\rm Im} f$; we will then define
${\rm Re} f\equiv \mu$ and call it {\it generalized moment map}. We ask the
reader to accept the name for the time being: it will be explained in the
next section.

This discussion, however, could be repeated {\sl verbatim} for the $-$ sector,
leading, at this stage, to {\it two} moment maps $\mu_\pm$. We will see
shortly that gauging the invariance under $\xi$ requires these two to be
equal.

\subsection{Gauging: review of the bosonic case}
On our way to gauging (\ref{eq:ungauged}), we now review quickly the bosonic
case, following \cite{hs,jjmo}. The bosonic action reads
\begin{equation}
    \label{eq:bosung}
    S= -\frac12\int_\Sigma g_{mn}d\phi^m\wedge* d\phi^n +\int_B H\ ;
\end{equation}
again we call $\xi$ the vector under which the model is invariant,
which means that $\xi$ is an isometry ($L_\xi g=0$) and that
$L_\xi H=0$. As always, although the action contains an integral
over the three--dimensional manifold $B$, the equations of motion
are two-dimensional, because a field variation $L_{\delta\phi}$
acts as follows:
\begin{equation}
    \label{eq:stokes}
L_{\delta\phi} \int_B H= \int_B \{d , \iota_{\delta\phi}\}H =
\int_B d(\iota_{\delta\phi} H) = \int_\Sigma \iota_{\delta\phi} H
\end{equation}
where we have used the ``magic Cartan formula''
\begin{equation}
    \label{eq:cartan}
    L_v = d \iota_v + \iota_v d
\end{equation}
valid for any vector field $v$.

Gauging the model means that we want to promote invariance under $\xi$ to
an invariance
\begin{equation}
    \label{eq:lav}
\delta \phi = \lambda \xi(\phi) \ ,
\end{equation}
with $\lambda$ a function on $\Sigma$. This is accomplished by
introducing a vector field $A$ which transforms as
\begin{equation}
    \label{eq:laA}
 \delta A = - d\lambda\ .
\end{equation}
To write the gauged action, it is convenient to introduce the
covariant derivative $d^A= d\phi + A \xi$, so that $d^A (f(\phi))=
df(\phi)+ A \xi^m \frac{\del f(\phi) }{\del\phi^m}$ . It is
covariant in the sense that
\[
\delta(d^A\phi)=\lambda d^A(\xi)\ .
\]
One is familiar with various particular cases of this covariant derivatives,
notably the case in which the components of $\xi$ are linear in $\phi$ as in most
gauge theories, and the case in which they are constant, which appears for
example in many gauged supergravity theories.

One is tempted to change all the derivatives $d$ in the action
(\ref{eq:bosung}) into covariant $d^A$'s; this, however, would
spoil the Stokes argument in (\ref{eq:stokes}). One way around
this is to introduce a compensating three--dimensional integral.
The other way, which we will use in this paper, is to write the
gauged action as
\begin{equation}
    \label{eq:bosgau}
    S= -\int_\Sigma \left(\frac12 g_{mn}d^A\phi^m\wedge* d^A\phi^n + \alpha \wedge A +
    \frac{1}{e^2} dA\wedge * dA\right)+\int_B H
\end{equation}
with $e$ being the gauge coupling and $\alpha$ defined by
(\ref{eq:alpha}) (from now on we will drop the subscript $\xi$ on
$\alpha$). When varying the last term in parentheses in
(\ref{eq:bosgau}) with respect to $\phi$, one now gets
\[
\int_\Sigma \{\iota_{\delta\phi}, d\} \alpha A =\int_\Sigma\left(
\iota_{\delta\phi}\iota_\xi H A - \iota_{\delta\phi} \alpha
dA\right)\ ;
\]
the first term completes the variation of $H$ in (\ref{eq:stokes}) so that one
gets
\[
\delta\phi^m \, H_{mnp}d^A\phi^n \wedge d^A\phi^p
\]
from their combination. For the action to be invariant under the local
transformations (\ref{eq:lav}), (\ref{eq:laA}), it actually also turns
out \cite{hs} that the condition
\begin{equation}
    \label{eq:xia}
    \iota_\xi \alpha=0
\end{equation}
has to be satisfied. We will find it later in the supersymmetric case.

It is interesting to consider what happens if one follows the
renormalization group flow in this model. As in \cite{w-phases},
the gauge coupling $e$ diverges in the infrared limit, and hence
the kinetic term is negligible. This makes the gauge field $A$ an
auxiliary field, and one can integrate it out. If one does that,
one gets a sigma model with target $M'=M/{\rm U(1)}$, where the
U(1) action is generated by $\xi$. The metric $g'$ and the NS
three--form $H'$ on the manifold $M'$ are given  by
\begin{equation}
    \label{eq:redgB}
     g'_{mn}= g_{mn}-\xi^2 \tilde\xi_m \tilde \xi_n+\xi^{-2}\alpha_m \alpha_n\ ,
\qquad \tilde H' = H + d( \tilde \xi \wedge \alpha)\
\end{equation}
where
\[
\tilde\xi_m=\frac1{\xi^2} g_{mn}\xi^n
\]
is the one--form dual to $\xi$ (so that $\iota_\xi \tilde\xi=1$)
and $\xi^2= g_{mn} \xi^m \xi^n$. For $\alpha=0$, the form of the
metric in (\ref{eq:redgB}) would be the metric obtained on the
quotient by the Kaluza--Klein procedure. The extra piece is not
inconsistent, however, since the metric still satisfies $\xi^m
g'_{mn}=0$ thanks to (\ref{eq:xia}), hence it is well--defined for
tangent vectors on the quotient manifold (which are defined as
equivalence classes of tangent vectors on $M$ under the
equivalence relation $v\sim v+ c \xi, c\in\rr$). To understand
better the geometric meaning of (\ref{eq:redgB}), let us rewrite
$g'$ as
\begin{equation}
    \label{eq:alphapr}
  g'_{mn}=
g_{pq} Q_+^p{}_m Q_+^q{}_n = g_{pq} Q_-^p{}_m Q_-^q{}_n\ , \quad
{\rm where } \quad Q_\pm^m{}_n\equiv
\delta^m{}_n-\xi^m\Big(\tilde\xi_n\pm \frac{\alpha}{\xi^2}\Big)\ .
\end{equation}
The matrices $Q_\pm$ are projectors, i.e. $Q_\pm Q_\pm=Q_\pm$.
Their kernel is one-dimensional and consists of vectors
proportional to $\xi$. The image consists of vectors $v$
satisfying the condition $g_{mn}v^m \xi^n\pm \alpha(v)=0$. It is
easy to see that in any equivalence class of tangent vectors under
the relation $v\sim v+c\xi$ there is a unique representative which
belongs to the image of $Q_+$ (or $Q_-$). Thus the formula for the
metric $g'$ can be described in words as follows. Consider two
vectors $v'_{1,2}$ on the quotient manifold $M'$, whose scalar
product we want to compute. We can think of them as tangent
vectors on $M$ defined up to shifts by $\xi$. In each of the two
equivalence classes we find the (unique) representatives $v_1$ and
$v_2$ which satisfy $\langle v_i,\xi\rangle+\alpha(v_i)=0$ (we
chose to work with $Q_+$ for definiteness). We define the scalar
product of $v'_1$ and $v'_2$ to be the scalar product of $v_1$ and
$v_2$ with respect to the metric $g$. (In the KK procedure, one
would take representatives $v_{1,2}$ which are orthogonal to
$\xi$). This is what is expressed by (\ref{eq:alphapr}).

We will see below that the projectors $Q_\pm$ in
(\ref{eq:alphapr}) play a similar role for all tensors in the
gauged supersymmetric model.

Let us also notice that $H'$ in (\ref{eq:redgB}) is a basic form,
and hence it also lives on the quotient $M'$. By definition, a
basic form is annihilated both by $\iota_\xi$ and by $L_\xi$. For
the first of these, one has to use (\ref{eq:alpha}),
(\ref{eq:xia}) and the identity $\iota_\xi d(\tilde \xi)=0$ (which
can be seen most easily by taking coordinates adapted to the
fibration, in which $\xi=\del_\psi$ for an angular coordinate
$\psi$). $L_\xi H'=0$ then follows easily from $H'$ being closed
and from (\ref{eq:cartan}).

\subsection{The gauged supersymmetric action}

We will now introduce the supersymmetric model, much along the
lines of the bosonic gauged model described in the previous
subsection. In analogy to the introduction of $A$ there, we want
to introduce an $\mathcal{N}=1$ vector multiplet $\Gamma_\alpha$
to gauge (\ref{eq:ungauged}). Since later we want to require that
the action be $\mathcal{N}=2$-supersymmetric, we also introduce an
extra $\mathcal{N}=1$ scalar multiplet $S$, which together with
$\Gamma_\alpha$ should form an $\mathcal{N}=2$ vector multiplet.

All in all the action we introduce reads
\begin{eqnarray}
\label{eq:Sg} &&
    -\int_\Sigma g_{mn} D_+^\Gamma \phi^m D_-^\Gamma \phi^n\ d^2x d^2\theta +
\int_B  H_{mnp} \del_t \phi^m D_+ \phi^n  D_- \phi^p d^3x d^2\theta\\
\nn && +\int_\Sigma \left(\frac1{e^2} (W_\alpha W^\alpha+ D_\alpha
S \, D^\alpha S) -
 S \,\mu(\phi)- \alpha_m (D_+\phi^m \Gamma_-+ D_-\phi^m \Gamma_+
 )\right) d^2x d^2\theta
\end{eqnarray}
where the covariant derivative
\[
D_\alpha^\Gamma= D_\alpha+\Gamma_\alpha \xi^m \frac{\del}{\del \phi^m}\ ;
\]
for example $D_\alpha^\Gamma \phi^m = D_\alpha \phi^m + \Gamma_\alpha \xi^m$.

So far this is a (1,1) model. We now want to see what are the
consequences of imposing a second supersymmetry in both
left-moving and right-moving sectors, similarly to the result in
\cite{ghr} quoted around equation (\ref{eq:susy}).

\subsection{The second supersymmetry transformation}
\label{sec:second}
The second supersymmetry transformation for all the fields is easy to guess.
Let us start from the $\mathcal{N}=2$ vector, made up by $\Gamma_\alpha$
and $S$. If we had an $\mathcal{N}=2$ superfield $V$, we could expand its
dependence on $\theta^2_\alpha$ as
\[
V=v-i\theta_2^{\,\alpha}\, \Gamma_\alpha + i\theta_2^+\theta_2^- S
\]
where $v,\Gamma_\alpha,S$ are functions of  $\theta_1$ only, and the
factors are for later convenience. The
gauge transformation for such a superfield consists of shifting
$V\to V+{\rm Re}\Lambda$, with $\Lambda$ a chiral $\mathcal{N}=2$ superfield.
In particular, we can use this gauge freedom to send $V$ to an ``$\mathcal{N}=1$
Wess--Zumino gauge'' in which $v=0$, by choosing $\Lambda$ so that
$\Lambda_{|_{\,\theta^\alpha_2=0}}=v/2$. This gives
\begin{equation}
    \label{eq:wz}
V\to -i\theta_2^\alpha \Gamma_\alpha + \theta_2^+\theta_2^-(iS-D^2 v)
\end{equation}
In $\mathcal{N}=2$ terms, the second supersymmetry transformation is
easy to express as $\delta_2 V=
\epsilon^\alpha D_{2\,\alpha} V_{|_{\,\theta^\alpha_2=0}}$. This gives
the transformation laws
\begin{equation}
    \label{eq:deltaV}
\delta_2(v,\Gamma_\alpha,S)=-\epsilon_2^\beta(i\Gamma_\beta, \,
 (\del_{\alpha\beta} v- \epsilon_{\alpha\beta}S),\,
-i\del_{\alpha\beta} \Gamma^\alpha)
\end{equation}
Now, by starting from $(0,\Gamma_\alpha, S)$ and composing (\ref{eq:deltaV})
and (\ref{eq:wz}), one gets
\begin{equation}
    \label{eq:deltaSG}
    \delta \Gamma_\alpha = \epsilon_\alpha S \ , \qquad
\delta S= \epsilon^\alpha W_\alpha\
\end{equation}
where we have defined
\begin{equation}
    \label{eq:W}
    W_\alpha= D^\beta D_\alpha \Gamma_\beta\ .
\end{equation}

As for the $\phi^m$, we will generalize in the simplest way the
supersymmetry transformations in (\ref{eq:ung-deltaphi}):
\begin{equation}
    \label{eq:deltaphi}
\delta \phi^m= \epsilon^+ D_+^\Gamma \phi^n I^m_{+\,n}  +
\epsilon^- D_-^\Gamma \phi^n I^m_{-\,n} \ .
\end{equation}
This is in fact the only possible expression for the second
supersymmetry transformation which is gauge-invariant and
compatible with dimensional analysis.

It is not very difficult to check that the putative second
supersymmetry transformation in (\ref{eq:deltaphi}) commutes with
the one implicit in the superfield notation; it follows in a
standard way from the fact that $D$ and $Q$ commute. Much less
trivial is the fact that the second supersymmetry transformation
obeys by itself the supersymmetry algebra. But first we will
require that it leaves our action invariant.

\subsection{Invariance of the action}
\label{sec:inv} First of all, one can check that supersymmetry
variations of the $(D_\alpha S)^2$ and $(W_\alpha)^2$ terms in
(\ref{eq:Sg}) annihilate each other. To check this, one needs to
use the identity $D_\alpha W_\beta=D_\beta W_\alpha$ which follows
from $D_\alpha W^\alpha=0$, which in turn follows from
(\ref{eq:DD}).)

The other terms contain $\phi$ and are much more involved. The
result is that for the action to be supersymmetric the following
has to be satisfied:
\begin{eqnarray}
    \label{eq:invactionold}
&&(\omega^\pm)^t=-\omega^\pm \ , \qquad \nabla_\pm \omega_\pm=0 \ ; \\
&& \label{eq:invactionnew}
   L_\xi g=L_\xi \omega=0\ , \qquad
\omega_\pm \xi \mp I^t_\pm \alpha= d\mu\ , \qquad \iota_\xi \alpha=0
\end{eqnarray}
where we recall that $\omega^\pm_{mn}\equiv g_{mp}I^p_{\pm\,n}$. We have also
used $(\omega \xi)_m=\omega_{mn}\xi^n$ for $-(\iota_\xi \omega)_m$.
Notice that the conditions in the first line, (\ref{eq:invactionold}),
were already present for the ungauged model, as mentioned in section
\ref{sec:ungauged}; the second line, (\ref{eq:invactionnew}), contains the
conditions specifically arising upon gauging.

Actually, by looking at the second equation in
(\ref{eq:invactionnew}), we recognize that it was almost implied
(locally) by the invariance of the ungauged model; the difference
is that, at that point, one could only derive the existence of two
separate moment maps. What we are finding here is that to write
down a supersymmetric gauged model those two moment maps have to
be equal. This can be traced back to the fact that in the gauged
action (\ref{eq:Sg}) there is only room for one function
$\mu(\phi)$, and not for two.

\subsection{Supersymmetry algebra}
\label{sec:sualg} We now require that the transformations defined
in section \ref{sec:second} satisfy the right supersymmetry
algebra on-shell.

In fact, they only do so {\it up to gauge transformations}. Namely, in appendix
\ref{app:boring} we show that (\ref{eq:deltaV}), (\ref{eq:deltaSG}) and
(\ref{eq:deltaphi}) satisfy
\begin{eqnarray}
\nn&&
[\delta_1,\delta_2]S=-2i(\epsilon_1^+\epsilon_2^+ \del_{\+}
+\epsilon_1^-\epsilon_2^- \del_=) S \ , \\
&&[\delta_1, \delta_2 ]\Gamma_\alpha
=-2i(\epsilon_1^+\epsilon_2^+ \del_{\+} +
\epsilon_1^-\epsilon_2^- \del_=)\Gamma_\alpha + \Lambda\ ,
    \label{eq:susy-gauge}\\
\nn&&[\delta_1, \delta_2]\phi^m=-2i(\epsilon_1^+\epsilon_2^+ \del_{\+} +
\epsilon_1^-\epsilon_2^- \del_=)\phi^m -\Lambda\,\xi^m\ ,
\end{eqnarray}
with a gauge transformation $\Lambda$ that happens to be
\begin{equation}
    \label{eq:Lambda}
\Lambda=2\epsilon_1^+\epsilon_2^+ \, D_+\Gamma_+ + ( \epsilon^1_+ \epsilon^2_-
+ \epsilon^2_+ \epsilon^1_-) ( D_+\Gamma_- + D_-\Gamma_+)+
2\epsilon_1^-\epsilon_2^- \, D_- \Gamma_-
\end{equation}
if the following conditions are satisfied:
\begin{equation}
    \label{eq:susyalgebra}
    (I_\pm)^2=-1\ , \qquad {\rm Nij}(I_\pm)=0\ ; \qquad \qquad
L_\xi I_\pm=0\ ,
\end{equation}
${\rm Nij}$ being the Nijenhuis tensor.
This time, the only condition really specific to the gauged model is the last one,
whereas the first two (that $I_\pm$ are complex structures) already arise for
the ungauged model.

\subsection{The flow to the infrared}
As in the bosonic case, the renormalization group flow makes the
kinetic term for $\Gamma_\pm$ and $S$ in (\ref{eq:Sg}) negligible
and $\Gamma_\pm$ and $S$ non--dynamic. Integrating out
$\Gamma_\pm$ then gives
\begin{equation}
    \label{eq:G=}
\Gamma_\pm=-(\xi^n g_{mn}\pm \alpha_m)D_\pm \phi^m  \
\end{equation}
which will be useful later. At the same time $S$ becomes a
Lagrange-multiplier superfield which constrains the bosonic fields
to the hypersurface $\mu(\phi)=0$. Hence the infrared limit is an
ungauged sigma-model whose target $M'$ is the quotient of the
hypersurface $\mu=0$:
\begin{equation}
    \label{eq:GIT}
M'=\{\mu=0\}/{\rm U(1)}
\end{equation}
as familiar from \cite{w-phases}. The formul\ae\ for $g'$ and $H'$
in (\ref{eq:alpha}) are still valid, with a similar
interpretation; the only extra step is that the vectors $v'$ in
the discussion after (\ref{eq:alphapr}) have to be tangent to $\{
\mu=0\}$.

It is instructive to see what happens to the second supersymmetry
transformation in the infrared, when we integrate out the fields
$\Gamma$ and $S$ as in (\ref{eq:G=}). One gets the usual
transformations in (\ref{eq:ung-deltaphi}), with complex
structures
\begin{equation}
    \label{eq:redIpm}
    I_\pm^{'\,m}{}_n =I_\pm^m{}_p Q_\pm^p{}_n\
\end{equation}
with $Q_\pm$ given in (\ref{eq:alphapr}). The geometric meaning of
these formulas is as follows. Consider a tangent vector $v'$ on
the quotient $M'$ given by (\ref{eq:GIT}). Again since $Q_\pm
\xi=0$, this can be thought of as a vector on $\{ \mu=0\}$ defined
modulo $v'\to v'+\lambda\xi$. There is a unique representative in
the equivalence class such that $g_{mn}\xi^m v^n\pm \alpha(v)=0$.
Now one can act on this representative with $I_\pm$ to get a new
vector on $M$; one can verify that this vector is tangent to
$\{\mu=0\}$ by computing $\del_m \mu\, I_\pm^{'\,m}{}_n=0$. This
is done by noticing that $Q_\pm$ in (\ref{eq:alphapr}) can be
rewritten, using the second equation in (\ref{eq:invactionnew}),
as
\[
Q_\pm^m{}_n= \delta^m{}_n-\frac{\xi^m}{\xi^2}I_\pm^p{}_n\del_p \mu
\]
and by noticing that $\del_m \mu\, I_\pm^m{}_n \xi^n=\xi^2$ (using
this time both the second and third equation in
(\ref{eq:invactionnew})). One can now project the vector back to
the quotient $M'$; hence we have obtained a linear map from $TM'$
to itself, and this map is the complex structure on $M'$.

\section{Geometrical interpretation}
\label{geometry} We are now ready to reinterpret the conditions we
got for the existence of a gauged sigma model in terms of
generalized geometry. The ungauged analogue of this is the
equivalence  between bihermitian geometry and generalized K\"ahler
geometry proved in \cite{gualtieri}. Therefore we first give a
lightning review of this correspondence. For more details, see
\cite{gualtieri}.

\subsection{Bihermitian and generalized K\"ahler geometries}
\label{sec:rev}
An {\it almost generalized complex structure} $\cal J$ on a manifold is
an endomorphism of $T\oplus T^*$, squaring to -1, and hermitian with respect
to the metric
\[
\mathcal{I}=
\left(\begin{array}{cc}
0&1 \\1&0
\end{array}\right)
\]
 on $T\oplus T^*$.  It follows
that it has
eigenvalues $\pm i$; call them $L_\mathcal{J}$ and $\bar L_\mathcal{J}$. The
so--called {\it twisted Courant bracket} is the derived bracket \cite{yks} on
$T\oplus T^*$ with respect to the differential
$d+ H\wedge$, where $H$ is a closed three--form.
Now, if $L_\mathcal{J}$ is closed with respect to the twisted Courant bracket,
one says that $\mathcal{J}$ is {\it twisted integrable}.

If we have two commuting generalized complex structures
$\mathcal{J}_{1,2}$ such that their product $M\equiv{\cal I}
\mathcal{J}_1 \mathcal{J}_2$ is a positive definite metric on
$T\oplus T^*$, we say that the manifold is {\it generalized
K\"ahler}.

The reason this geometry is relevant for us is that such a
pair $\mathcal{J}_{1,2}$ can be shown to have the form
\begin{equation}
    \label{eq:gk}
\mathcal{J}_{1,2}=
\frac12    \left(\begin{array}{cc}
I_+ \pm I_- & - (\omega^{-1}_+ \mp \omega^{-1}_-)\\
\omega_+ \mp \omega_- & -(I_+ \pm I_-)^t
    \end{array}\right)
\end{equation}
for some bihermitian structure defined by complex structures
$I^m_{+\,n}$ and two--forms $\omega_\pm$; in particular, twisted
integrability of $\mathcal{J}_{1,2}$ is equivalent to the
integrability of $I_\pm$  and $d\omega_\pm= \pm\iota_{I_\pm} H$
(which we had in (\ref{eq:susy})).

Alternatively, locally one can replace twisted integrability with
respect to $H$ with ordinary integrability, and replace
$\mathcal{J}_{1,2}$ with their so--called $b$--transform:
\[
\mathcal{J}_{1,2} \to
\left(\begin{array}{cc}
1& 0 \\ b & 1
\end{array}\right)
\mathcal{J}_{1,2}
\left(\begin{array}{cc}
1& 0 \\ -b & 1
\end{array}\right) \ .
\]

The K\"ahler case is recovered in these formul\ae\ by taking $H=0$, $I_+=I_-$
and $\omega_+=\omega_-$.


Finally, let us mention a construction that we will need in
section \ref{sec:examples}. To any generalized complex structure
$\mathcal{J}$ one can associate locally an inhomogeneous
differential form $\Phi$ (called {\it pure spinor}) with certain
special properties . There are two features of this correspondence
that we will need later.

The first is that twisted integrability for $\mathcal{J}$
translates into the existence of a one--form $\eta$ and a vector
field $v$ such that $(d+H\wedge)\Phi=(\eta\wedge+ \iota_v)\Phi$.

The second one concerns the {\it type } of a pure spinor $\Phi$.
This is defined as the smallest degree of a homogeneous component
of $\Phi$. It can be shown \cite{gualtieri} that the type of a
pure spinor $\Phi$ is equal to the number of $i$--eigenvectors of
the corresponding $\mathcal{J}$ of the form $(v,0)^t$ -- that is,
the dimension of the intersection of the $i$--eigenspace of
$\mathcal{J}$ with $T$.

For more details, again see
\cite{gualtieri,hitchin}.

\subsection{Generalized moment map}
After having reviewed how the conditions from the ungauged sigma model
can be cast in the language of generalized K\"ahler geometry, we will now
look at the conditions coming from the {\it gauged} (2,2) sigma model.

First we recall what an ordinary moment map is. If a vector
preserves a symplectic form $\omega$ ($L_\xi \omega=0$), one has
by (\ref{eq:cartan}) that $d\iota_\xi \omega=0$. Then locally one
has
\begin{equation}
    \label{eq:usualmm}
    \iota_\xi \omega=-d\mu
\end{equation}
for some function $\mu$. This function is called the moment map.

Consider now the second equation in (\ref{eq:invactionnew}):
\begin{equation}
    \label{eq:inv-mom}
     \omega_\pm\xi \mp I^t_\pm \alpha= d\mu \ .
\end{equation}
If one takes sum and difference of these equations, one gets
\begin{equation}
    \label{eq:Jmom}
    {0\choose d\mu}=
\frac12    \left(\begin{array}{cc}
I_+ - I_- & - (\omega^{-1}_+ + \omega^{-1}_-)\\
\omega_+ + \omega_- & -(I_+ -I_-)^t
    \end{array}\right)
{\xi \choose \alpha_\xi} =
{\cal J} {\xi \choose \alpha_\xi} \ .
\end{equation}
(In the notation of \cite{gualtieri}, ${\cal J}={\cal J}_2$.)
In other words, ${\cal J}(\xi+ \alpha)= d\mu$.
This is equivalent to ${\cal J} (\xi+ \alpha - i d\mu)=
i(\xi+ \alpha - i d\mu)$, or in other words
\begin{equation}
    \label{eq:lt2}
    \xi+ \alpha - i d\mu \in L_{\cal J}\ .
\end{equation}
This is exactly the definition of a {\it generalized moment map}
in \cite{lt2}. An action which admits a generalized moment map is
called a generalized hamiltonian action.

This name is well motivated: in the K\"ahler case, $I_+=I_-$ and
 $\mathcal{J}$ in (\ref{eq:Jmom}) becomes
\[
\left(
    \begin{array}{cc}
0 & -\omega^{-1}\\ \omega & 0
    \end{array}\right)
\]
so that $\alpha=0$ and $d\mu=\omega \xi$, just as in (\ref{eq:usualmm}).

We would like to emphasize that the existence of the generalized
moment map puts a strong additional constraint on the vector field
$\xi$ and the corresponding one-form $\alpha$, even assuming that
$\xi$ preserves all the tensors involved. This is in contrast with
the K\"ahler case, where the moment map always exists locally,
though there may be global obstructions coming from the nontrivial
topology of the target manifold. In the case when $H\neq 0$, if we
wanted to gauge while preserving only ${\mathcal N}=(2,1)$
supersymmetry, the situation would be similar: locally we can
always solve equation (\ref{eq:del}) for $f$, while globally we
may find an obstruction living in the Dolbeault cohomology group
$H^1_\partial(M)$. But in the ${\mathcal N}=(2,2)$ case we find an
extra strong constraint coming from the requirement that
right-moving and left-moving moment maps be identical. We have
shown above that this physical constraint corresponds to the
requirement that the action by $\xi$ be generalized hamiltonian in
the sense of \cite{lt2}.

Another condition in (\ref{eq:invactionnew}) was that $\iota_\xi
\alpha=0$. From (\ref{eq:lt2}), since $L_\mathcal{J}$ is isotropic
(that is, the metric $\mathcal{I}= {{0\ 1}\choose {1\ 0}}$ is zero
when restricted to $L_\mathcal{J}$), we know that
$\iota_\xi(\alpha-id\mu)=0$. Taking the real part, we obtain the
desired relation, which therefore is not independent. If we had
considered a quotient by a group $H$ more complicated than U$(1)$,
this relation would have been non--trivial (it would have
corresponded to the condition that the moment map be {\it
equivariant} in \cite{lt2}).

As for
the remaining conditions in (\ref{eq:invactionnew}), (\ref{eq:susyalgebra}),
together they say
\begin{equation}
    \label{eq:preserve}
L_\xi \mathcal{J}_{1,2}=0 \ .
\end{equation}

Hence we have now reinterpreted all the conditions for on--shell
(2,2) supersymmetry of the gauged sigma model in terms of
generalized K\"ahler geometry. To summarize:
\begin{itemize}
  \item the manifold has to be generalized K\"ahler (just like
in the ungauged case);
\item $\xi$ has to preserve the generalized K\"ahler structure (eq.(\ref{eq:preserve}));
\item a generalized moment map $\xi$ (with one--form $\alpha$) has to exist for
the action of $\xi$.
\end{itemize}

Looking at \cite{lt2}, we find that these are precisely the
hypotheses for their Proposition 12. Hence the quotient by $\xi$
inherits a twisted generalized K\"ahler structure. Above we have
determined the bihermitian structure on the quotient using
physical methods (integrating out the fields $\Gamma$ and $S$).
One possible way to read our result is as a physical proof of the
theorem in \cite{lt2}.

\section{Examples}
\label{sec:examples}

We give here a few very simple examples, taking inspiration from some of
the mathematical papers on generalized K\"ahler reduction.

{\bf 1.} Perhaps the simplest twisted generalized K\"ahler
structure is the one living on the Hopf surface which is
topologically $S^3\times S^1$ \cite{rss,gualtieri}. This can be
described by $\cc^2-\{0\}$ with complex coordinates $z_1$, $z_2$,
quotiented by $z_i\to 2 z_i$.  The two complex structures $I_\pm$
give respectively $i,i$ and $i,-i$ on $\del_{z_1}$, $\del_{z_2}$,
and hence they commute. The three--form flux $H$ is the volume
form of $S^3$. If we let $z_1=r \sin\lambda\, e^{i\phi_1}, z_2=r
\cos\lambda\, e^{i\phi_2}$, where $\lambda\in
\left[0,\frac{\pi}{2}\right]$ and both $\phi_1$ and $\phi_2$ have
period $2\pi$, then the metric is
$$
ds^2=\frac{\sum_{i=1}^2 dz_i
d\bz_i}{r^2}=\frac{dr^2}{r^2}+d\lambda^2+\sin^2\lambda\,
d\phi_1^2+\cos^2\lambda\, d\phi_2^2.
$$
The three-form $H$ is
$$
H=\sin 2\lambda\, d\lambda d\phi_1 d\phi_2.
$$
If we do not quotient by $z_i\to 2 z_i$, we get the product metric
on $S^3\times \RR$. In the context of string theory, it describes
the near-horizon geometry of the Neveu-Schwarz fivebrane.

This geometry has $SU(2)\times SU(2)\times \RR$ isometry group,
where $SU(2)\times SU(2)$ acts on $S^3\simeq SU(2)$ by right and
left translations, while $\RR$ acts by $r\to r e^t$. One could try
to reduce this model either along one of the left--invariant
vector fields on $S^3$ or along $r\frac{\partial}{\partial r}$. It
turns out that this is not possible: even though these vector
fields preserve all the tensors concerned, their action is not
generalized hamiltonian -- that is, no $\mu$ exists so that
(\ref{eq:Jmom}) is satisfied. One can see this in the following
way. From (\ref{eq:inv-mom}) one can derive
\begin{equation}
    \label{eq:comm}
        [I_+,I_-] g^{-1}\alpha= (2+ \{I_+, I_-\}) \xi\ ,
\end{equation}
an equation that we will need again in the appendix. If the two complex structures
commute, the left hand side is zero; also, the right hand side then becomes
proportional to $P\xi\equiv (1/2)(1+I_+I_-)\xi$. Now, $I_+ I_-$ is an almost
product structure \cite{ghr}, and $P$ is a projector defined by it; so we have
found that $\xi$ is along one of the two subspaces defined by the almost
product structure. This is a general result. Its physical interpretation is
that, when the model can be written without using semi--chiral multiplets, the
gauging may involve either only the chiral or only the twisted chiral multiplets.

Coming back to $S^3\times \RR$, from the explicit form of $I_\pm$
given above one sees that $\xi$ should either involve only $z_1$
or only $z_2$. Of course the two possibilities are equivalent, so
let us pick the first one and write $\xi=iz_1\del_{z_1}-i \bar
z_1\del_{\bar z_1}$. The reduction along this vector field has
been considered by S.~Hu \cite{hu}. Let us describe this example
in some detail; below we modify it to produce a new family of
generalized K\"ahler structures on $S^3\times \RR$ and $S^3\times
S^1$. First we need to choose a one-form $\alpha$ which solves the
equation $i_\xi H=d\alpha$. The choice which leads to a
hamiltonian action turns out to be
$$
\alpha=\cos^2\lambda d\phi_2.
$$
Note that this form is smooth everywhere on $S^3$. The corresponding moment map turns out to be
$$
\mu=-\log r +const.
$$
Thus the zero-level of the moment map is a submanifold given by $r=const$, which is a three-sphere.
The quotient of this submanifold by the vector field
$$
\xi=\frac{\partial}{\partial\phi_1}
$$
can be parametrized by $\lambda$ and $\phi_2$ and can be identified with a disc. The reduced metric turns out to be
$$
d\lambda^2+\tan^2\lambda d\phi_2^2.
$$
This is precisely the metric which corresponds to the ${\mathcal
N}=2$ minimal model $SU(2)/U(1)$ \cite{MMS}. This is hardly
surprising: the first step in the generalized K\"ahler reduction
in this case amounts to fixing $r$ to be constant, thereby
reducing the theory to the ${\mathcal N}=(1,1)$ $SU(2)$ WZW model,
while the second step consists of gauging the adjoint action of
the maximal torus of $SU(2)$ and integrating out the gauge
${\mathcal N}=(1,1)$ supermultiplet, which gives the supercoset
$SU(2)/U(1)$.

We can modify the above construction to produce a one-parameter
family of generalized K\"ahler structures on $S^3\times \RR$ and
$S^3\times S^1$. Consider $S^3\times \RR\times \RR^2$, where we
regard $\RR^2$ as a flat K\"ahler manifold with a complex
coordinate $z_3=x_3+iy_3$ and a metric $ds^2=|dz_3|^2$. Let us
quotient this generalized K\"ahler manifold by a vector field
$$
iz_1 \frac{\partial}{\partial
z^1}-i\bz_1\frac{\partial}{\partial\bz_1}+\zeta\frac{\partial}{\partial
y_3},
$$
where $\zeta$ is an arbitrary real number. (More generally, we
could consider a product $S^3\times \RR\times Y$, where $Y$ is a
K\"ahler manifold with a $U(1)$ symmetry). The moment map is now
$$
\mu=-\log r - \zeta x_3+const.
$$
The equation $\mu=0$ allows to express $x_3$ in terms of $r$, so
the quotient of the submanifold $\mu=0$ can be naturally
identified with $S^3\times \RR$ parametrized by
$r,\lambda,\phi_1,\phi_2$. The reduced metric is
$$
\frac{dr^2}{r^2}\left(1+\frac{1}{\zeta^2}\right)+d\lambda^2+\frac{\zeta^2\sin^2\lambda
d\phi_1^2+ \cos^2\lambda (1+\zeta^2)
d\phi_2^2}{\sin^2\lambda+\zeta^2}.
$$
In the limit $\zeta\to\infty$ it reduces to the standard metric on
$S^3\times \RR$. The three-form $H'$ on the reduced manifold is
$$
H'=\sin2\lambda\, d\lambda d\phi_1
d\phi_2+d(\tilde\xi\wedge\alpha)=\sin2\lambda\, d\lambda d\phi_1
d\phi_2+d\left(\frac{\sin^2\lambda\cos^2\lambda\, d\phi_1
d\phi_2}{\sin^2\lambda+\zeta^2}\right).
$$

Note that the reduced metric is invariant with respect to $r\to r
e^t$, so we can make periodic identification of $\log r$ and
produce a one-parameter deformation of the standard generalized
K\"ahler structure on $S^3\times S^1$. We can compute the
corresponding forms $\omega'_\pm$ following the same geometric
procedure as for the metric $g'$: we restrict $\omega_\pm$ to the
hypersurface $\mu=0$ and define the value of $\omega'_\pm$ on the
vectors $v'_{1,2}$ tangent to the quotient to be the value of
$\omega_\pm$ on the specially chosen representatives of $v'_{1,2}$
(those which lie in the image of $Q_\pm$). In this way we obtain:
$$
\omega'_+=-\frac{1+\zeta^2}{\sin^2\lambda+\zeta^2}\left[\sin\lambda\cos\lambda
\left(
 \frac{\zeta^2}{1+\zeta^2}d\lambda d\phi_1- d\lambda
d\phi_2\right)+\sin^2\lambda \frac{dr d\phi_1}{r}+\cos^2\lambda
\frac{dr d\phi_2}{r}\right]\ .
$$
One can also compute the complex structure $I'_+$; by
finding the $(1,0)$ forms with respect to this complex structure and
integrating them, one obtains
complex coordinates $z'_{1+}=r^{1+\zeta^{-2}} \sin\lambda \, e^{i\phi_1}$,
$z'_{2+}= r \cos\lambda \, e^{i\phi_2}=z_2$.
The form $\omega'_-$ and the complex structure $I'_-$ are obtained from
$\omega'_+$ and $I'_+$ by changing $\phi_2\to -\phi_2$.

{\bf 2.} We now want to give an example without NS flux, but with
non--commuting
complex structures. For this, we turn to \cite{bcg,lt1}. These authors apply
to $\cc^k$ a procedure devised in \cite{gualtieri} to deform an ordinary
K\"ahler structure into a generalized K\"ahler one. To describe the idea we
will need the pure spinors $\Phi$ introduced in section \ref{sec:rev}.
The pure spinors for the initial K\"ahler case corresponding
to the generalized complex structures (\ref{eq:gk}) read $\Phi_1=\Omega$, $\Phi_2=
e^{i\omega}$, where $\Omega$ is the $(k,0)$ form (it would in general only
exist locally, but we are considering $\cc^k$) and $\omega$ is the K\"ahler form.
Now the deformation is described by
\[
\Phi_1\to \exp[\beta^{ij} (\del_i-i\omega_{i\bar k} d\bar z^{\bar
k}) (\del_j+i\omega_{j\bar l} d\bar z^{\bar l})]\Phi_1\ ,
\]
where $\beta$ is a holomorphic Poisson bivector.
It so happens that the same operator acting on $\Phi_2$ leaves it invariant.

Choosing different bivectors $\beta$ and reducing along different
vector fields $\xi$ produces many examples of generalized K\"ahler
structures on a certain class of toric manifolds (not on all,
because of the condition that the bivector $\beta$ be holomorphic
Poisson). Unfortunately, it appears that for all these examples
the generalized K\"ahler structure before reduction is defined not
on all of $\cc^k$, but on some open set obtained by excluding a
lower-dimensional submanifold (defined by a real equation). For
instance, the examples in \cite{bcg} start with $\cc^3$ and reduce
it by the action $z^i\to e^{i\psi} z^i$, to yield a generalized
K\"ahler structure on $\cc\pp^2$. In order for the generalized
K\"ahler structure to be invariant, one has to take the Poisson
bivector to have degree two in the $z^i$, so that $\Phi_1$ is
homogeneous of degree 3. The Poisson bivector being non--constant
causes the norm of $\Phi_1$ go to zero on a  certain locus and the
generalized K\"ahler is not well--defined there.

Mathematically this is harmless, since one can usually arrange so
that the hypersurface $\mu={\rm const}$ does not intersect with
the troublesome locus. Physically, however, the model one starts
with has to be defined on the whole of the manifold in order for
the gauged sigma-model to make sense.

A way to circumvent this problem is as follows. Let us start from $\cc^4$ and
take the (integrated) action of $\xi$ to be
$(z^1,z^2,z^3,z^4)\to (e^{i\psi}z^1,e^{i\psi}z^2, e^{-i\psi}z^3,e^{-i\psi}z^4)$.
Then the pure spinors we start with are both of degree zero with respect to the
action of $\xi$. This means that we can take the bivector $\beta$ to be constant.
This does not lead to the problem described above, and one can safely perform
the reduction, getting a bihermitian structure on the conifold.

Unfortunately, by taking $\beta$ to be constant we have given up
the NS flux as well. Also, it should  be emphasized that we have not
changed the flat metric on $\cc^4$: the model we start with is still
the usual free sigma model, only with a very particular choice of a (2,2)
supersymmetry algebra. What one produces after reduction is a pair
of non--commuting, complex structures on the conifold,
covariantly constant with respect to the Levi--Civita connection given by
the reduced metric. This metric has holonomy $U(3)$; the Calabi--Yau metric
on the conifold is more complicated and is found by following the renormalization
group further down in the infrared.

So this bihermitian structure is not very
interesting {\sl per se}; it is, however, an example in which the
two complex structures $I_\pm$ do not commute. A way to see it is
the following. If $I_\pm$ commute, they are simultaneously
diagonalizable; by looking at (\ref{eq:gk}), we can produce $k$
eigenvectors of either $\mathcal{J}_1$ or $\mathcal{J}_2$ that are
purely in $T$. In other words, the sum of the types (see end of
section \ref{sec:rev})) of the two pure spinors $\Phi_1$ and
$\Phi_2$ is $k$.

For example, in the K\"ahler case, $\Phi_1=\Omega$ is a $k$--form
and has degree zero; $\Phi_1=e^{i\omega}=1+i\omega+\ldots$ has
inside the differential form 1, which has degree zero, so the type
of $\Phi_2$ is zero. The sum of the two is $k$, and indeed in this
case $I_+=I_-$. After the deformation by the bivector $\beta$,
however, the type of $\Phi_-$ is lowered; the sum of the two types
can no longer be zero, and by the reasoning above this means
$[I_+,I_-]\neq 0$.

\medskip

There are other constructions of generalized K\"ahler manifolds, but it is not
obvious whether they admit a hamiltonian action. Notably, the construction by
Hitchin \cite{hitchin-delpezzo}, that closely parallels the physical construction
in \cite{lrvz}, appears to be fairly general; and the even--dimensional semi--simple
groups are bihermitian, as pointed out in \cite{gualtieri}. It would be
interesting to consider their reduction.

\bigskip

{\bf Acknowledgments.} We would like to thank Marco Gualtieri for
useful correspondence. A.K. is supported in part by the DOE under
contract DOE-FG03-92-ER40701. A.~T. is supported by the DOE under
contract DEAC03-76SF00515 and by the NSF under contract 9870115.

\appendix
\section{Details about supersymmetry}
\label{app:boring}

We raise and lower indices with the tensor $\epsilon^{\alpha\beta}$, defined
so that $\epsilon^{+-}=1$.
The derivatives $D_\pm$ satisfy $(D_+)^2=i\del_\+$, $(D_-)^2=i\del_=$.
Some useful equalities are
\begin{equation}
    \label{eq:DD}
    D_\alpha D_\beta= i\del_{\alpha\beta}- \epsilon_{\alpha\beta}D^2\ ; \quad
D^2 D_\alpha= -D_\alpha D^2 = -i\del_{\alpha\beta} D^\beta \ ; \quad D^\alpha
D_\beta D_\alpha=0\
\end{equation}
where $D^2=D_+ D_-$.

Let us now look at the details of the computations in sections \ref{sec:inv}
and \ref{sec:sualg}. The methods are standard, if a little complicated; here
we list some of the steps in which the computation departs from the one for
the ungauged model.

For the variation of the action, it is useful to notice that
\[
\hspace{-.5cm}\frac{\delta}{\delta\phi^m}\left(\int g_{np} D^\Gamma_+ \phi^n D^\Gamma_- \phi^p
\right)=
-g_{mn} [D_+^\Gamma, D_-^\Gamma]\phi^n - 2 \Gamma_{mnp} D^\Gamma_+ \phi^n
D^\Gamma_- \phi^p + (\Gamma_+ D_-^\Gamma \phi^n-\Gamma_- D_+^\Gamma \phi^n)
(L_\xi g)_{mn}\ .
\]
(In computing this, one needs $\xi^m\del_m (D_+ \phi^n)=D_+\xi^n$.) This
can then be specialized to the variation under supersymmetry. Another slight
modification is given by the integration by parts. In the case of linear gaugings,
for example, all the covariant derivatives are given in the appropriate
representation, so that any scalar is acted on by a straight derivative; hence
one can integrate covariant derivatives by parts. In the present case, however,
a function may still be transforming non--trivially under the vector $\xi$.
One can, however, integrate by parts the straight derivative, and add and subtract
the connection piece, so that for example
\[
\int A_{[mn]} D_-^\Gamma D_+^\Gamma \phi^m D_+^\Gamma\phi^n =
\int\Big(-\frac12 A_{mn,p} D_-^\Gamma \phi^p  D_+^\Gamma \phi^m D_+^\Gamma \phi^n
+ \Gamma_- D_+^\Gamma \phi^m D_+^\Gamma \phi^n (L_\xi A)_{mn}\Big) \ .
\]
Using all this, the total variation is
\begin{eqnarray*}
&&\delta S= \int\epsilon^+\Big[
2 D^\Gamma_- D^\Gamma_+\phi^m D_+^\Gamma\phi^n \Big(-\omega^+_{(mn)}\Big)+
D_-^\Gamma \phi^p  D_+^\Gamma \phi^m D_+^\Gamma \phi^n
\Big( \nabla^+_p\omega^+_{mn}\Big)+\\
&&(S D_+^\Gamma\phi^m + D_+^\Gamma \phi^n I_+^m{}_n
(D_+\Gamma_-+D_-\Gamma_+))
\Big(g_{mp}\, \xi^p+\alpha_m-I_+^p{}_m\del_p \,\mu\Big)
+ S\Gamma_+\Big( \alpha_m \xi^m\Big)\\
&& \Gamma_+D_+^\Gamma \phi^m D_-^\Gamma \phi^n
\Big(I_+^p{}_m (L_\xi g)_{pn}\Big) +
\Gamma_+D_-^\Gamma \phi^m D_-^\Gamma \phi^n
\Big(I_+^p{}_m (L_\xi g)_{pn} +(L_\xi \omega)_{mn}\Big) \Big]+\\
&&\epsilon^-\Big[
2 D^\Gamma_+ D^\Gamma_-\phi^m D_-^\Gamma\phi^n\Big(\omega^-_{(mn)}\Big)+
D_+^\Gamma \phi^p  D_-^\Gamma \phi^m D_-^\Gamma \phi^n
\Big(- \nabla^-_p\omega^-_{mn}\Big)+\\
&&(S D_-^\Gamma\phi^m - D_-^\Gamma \phi^n I_-^m{}_n
(D_+\Gamma_-+D_-\Gamma_+))
\Big(g_{mp}\, \xi^p-\alpha_m-I_-^p{}_m\del_p \,\mu\Big)
+ S\Gamma_-\Big(- \alpha_m \xi^m\Big)\\
&& \Gamma_+D_-^\Gamma \phi^m D_+^\Gamma \phi^n
\Big(-I_-^p{}_m (L_\xi g)_{pn}\Big) +
 \Gamma_-D_+^\Gamma \phi^m D_+^\Gamma \phi^n
\Big(I_-^p{}_m (L_\xi g)_{pn} +(L_\xi \omega)_{mn}\Big) \Big]\ .
\end{eqnarray*}

Let us now look at the commutator of two
 second supersymmetry transformations. The one on $S$ is uneventful.
For the one on $\Gamma_+$, one only needs to add and subtract a term
$\epsilon^+_1 \epsilon^+_2 D_+^2 \Gamma_+$; one piece goes towards building
the right supersymmetry algebra, the other goes to
the gauge transformation $\Lambda$ given in (\ref{eq:Lambda}).
The most complicated commutator is obviously the one evaluated on $\phi^m$.
One gets:
\begin{eqnarray}
\nn&&    [\delta_1,\delta_2]\phi^m= \\
\nn&&
2\epsilon^+_1\epsilon^+_2\Big[(I_+^2)^m{}_n \,i\del_{\+}\phi^n + (I_+)^m{}_n
(D_+\Gamma_+)\xi^n+
D_+^\Gamma \phi^n D_+^\Gamma \phi^p\Big( {\rm Nij}(I_+)^m{}_{np}\Big) -
\Gamma_+ D_+^\Gamma \phi^n\Big((L_\xi I_+)^m{}_n\Big)\Big]+ \\
\nn&&2\epsilon^-_1\epsilon^-_2\Big[(I_-^2)^m{}_n \,i\del_=\phi^n + (I_-)^m{}_n
(D_-\Gamma_-)\xi^n+
D_-^\Gamma \phi^n D_-^\Gamma \phi^p\Big( {\rm Nij}(I_-)^m{}_{np}\Big) -
\Gamma_- D_-^\Gamma \phi^n\Big((L_\xi I_-)^m{}_n\Big)\Big]+ \\
\label{eq:boring}&&+(\epsilon^+_1\epsilon^-_2+\epsilon^-_1\epsilon^+_2)\Big[
S(I_+ - I_-)^m{}_n\,\xi^n+ \frac12[I_+,I_-]^m{}_n [D_+^\Gamma,D_-^\Gamma]\phi^n
\\
\nn&&+D_+^\Gamma \phi^n D_-^\Gamma \phi^p
\Big(I_+^m{}_q I_-^q{}_{p,n}+
I_+^m{}_{n,q} I_-^q{}_p-
I_-^m{}_q I_+^q{}_{n,p} - I_-^m{}_{p,q}I_+^q{}_n
\Big)\\
\nn&&+\frac12\{I_+,I_-\}^m{}_n\xi^n(D_+\Gamma_-+D_-\Gamma_+)
+\Gamma_+ D_-^\Gamma \phi^n\Big( -(L_\xi I_-)^m{}_n\Big)
+\Gamma_- D_+^\Gamma \phi^n\Big( -(L_\xi I_+)^m{}_n\Big)
\end{eqnarray}
We can now use some of the conditions coming from the action to massage the
$(\epsilon^+_1\epsilon^-_2+\epsilon^-_1\epsilon^+_2)$ term in this result.
First of all, one can show
\begin{equation}
    \label{eq:myst}
    [I_+,I_-]^m{}_q (\Gamma+\frac12 g^{-1}H)^q{}_{np}=I_+^m{}_q I_-^q{}_{p,n}+
I_+^m{}_{n,q} I_-^q{}_p-
I_-^m{}_q I_+^q{}_{n,p} - I_-^m{}_{p,q}I_+^q{}_n\ ,
\end{equation}
which is already useful in the ungauged case \cite{ghr}. To derive this identity, note that since $I_\pm$ are
covariantly constant with respect to the connections $\Gamma_\pm$, we can express ordinary derivatives of
$I_\pm$ in terms of $I_\pm$ and $\Gamma_\pm$:
$$
I_\pm^m{}_{n,p}=\Gamma_\pm^q{}_{pn} I_\pm^n{}_q-\Gamma_\pm^m{}_{pq}I_\pm^q{}_n.
$$
Substituting this into the expression on the r.h.s. of (\ref{eq:myst}), using
$\Gamma_-^m{}_{np}=\Gamma_+^m{}_{pn}$, and collecting similar terms, we get the expression on the l.h.s. of
(\ref{eq:myst}).

Using that $L_\xi g=L_\xi \omega_\pm=0$, we also get that $L_\xi I_\pm=0$.
Finally, from (\ref{eq:inv-mom}) we can derive
\begin{equation}
    \label{eq:IIstuff}
    [I_+,I_-] g^{-1}\alpha= (2+ \{I_+, I_-\}) \xi\ ,
\qquad
[I_+,I_-]g^{-1}d\mu =-2(I_+-I_-)\xi \ .
\end{equation}
The first of these two has already been used in section \ref{sec:examples}.
So the $(\epsilon^+_1\epsilon^-_2+\epsilon^-_1\epsilon^+_2)$ term in
(\ref{eq:boring}) now reads
\begin{eqnarray*}
    && [I_+,I_-]^m{}_n\Big( \frac12 [D_+^\Gamma,D_-^\Gamma] \phi^n+
\Gamma_+^n{}_{pq}D_+^\Gamma\phi^p D_-^\Gamma \phi^q -\frac12 S g^{np}
\del_p \mu +\frac12 g^{np} \alpha_p (D_+ \Gamma_- + D_-\Gamma_+)\Big)\\
&&-(D_+\Gamma_-+D_-\Gamma_+)\xi^m\ .
\end{eqnarray*}
The first line is now proportional to the equation of motion for $\phi$; the
second is a piece of the gauge transformation $\Lambda$ we claimed
in (\ref{eq:Lambda}) -- the other pieces having already been obtained in
(\ref{eq:boring}). This completes the computation.

\end{document}